\documentstyle[psfig,conf_iap,]{article}
\begin{document}
\heading{%
%
Scalar fields and cosmological attractor solutions \\
%
} 
\par\medskip\noindent
\author{%
Francesca Rosati$^{1}$
}
\address{%
Dipartimento di Fisica and INFN, \\
Universit\`a di Padova, via Marzolo 8, 
       I--35131 Padova, ITALY.
}

\begin{abstract}
Results of a general study \cite{nnr} on the dynamics of cosmological
scalar fields with arbitrary potentials are presented. 
Exact and approximate attractor solutions are found, with
applications to quintessence, moduli stabilization and inflation.
\end{abstract}

\section{Introduction}
Rolling scalar fields are the leading actors of many cosmological phenomena. 
Their dynamics is governed by two main 
ingredients: the steepness of the potential $V(\phi)$ and the equation 
of state of the background fluid $\gamma \equiv 1 + p_B/\rho_B$ ({\it e.g.} 
$\gamma = 4/3$ for radiation and $\gamma = 1$ for matter).
Cosmological attractor solutions have been found and studied  by several 
authors for various classes of potentials.
The aim of this work is to extend the results of 
\cite{copeland,scalcosmo} to a generic potential $V(\phi)$, 
and give analytical support to some of the conclusions in \cite{macorra}.
We make no assumption on the shape of the potential and allow the attractor 
scalar equation of state to slowly vary in time.
The solutions found can be read as field-dependent corrections to 
the attractors studied in \cite{copeland}, where the exponential 
potential was discussed.
Applications can span from quintessence to moduli stabilisation and inflation.

\section{The framework}

Taking $8\pi G = \kappa^2$, the variables \cite{copeland}
\begin{eqnarray}
x \equiv \frac{\kappa \phi'}{\sqrt{6}} , \hspace{1cm} 
y^2 \equiv \frac{\kappa^2 V}{3 H^2},
\end{eqnarray}
can be defined (a prime denotes a derivative with respect to the logarithm
of the scale factor $a$, $N \equiv \ln a$, and $H$ is the Hubble parameter); 
then the effective equation of state for the scalar field at any point yields,
\begin{equation}
\gamma_{\phi} \equiv \frac{\rho_{\phi} + p_{\phi}}{\rho_{\phi}} = 
\frac{\dot{\phi}^2}{\dot{\phi}^2/2 + V} = \frac{2x^2}{x^2+y^2} ,
\end{equation}
(which is constrained between $0 \le \gamma_{\phi} \le 2$),
and the contribution of the scalar field to the total energy density is given by
\begin{equation}
\Omega_{\phi} \equiv \kappa^2 \frac{\rho_{\phi}}{3 H^2} = x^2 + y^2 \,  
\end{equation}
(which is bounded between $0 \le \Omega_\phi \le 1$).

The relevant equations for a spatially--flat Friedmann--Robertson--Walker 
Universe read:
\begin{eqnarray}
\label{dynamics}
x' &=& -3x + \lambda \sqrt{\frac{3}{2}}y^2 + \frac{3}{2}x~[2x^2 +
       \gamma(1-x^2-y^2)] \nonumber \,, \\
y' &=& -\lambda\sqrt{\frac{3}{2}}x y + \frac{3}{2}y~[2x^2 +
       \gamma(1-x^2-y^2)] \,, \\
\lambda' &=& -\sqrt{6} \lambda^2 (\Gamma -1)x \nonumber \, .
\end{eqnarray}
where we have introduced  (see also \cite{macorra} and \cite{stein}) 
the important parameters
\begin{eqnarray}
\lambda \equiv - \frac{1}{\kappa V} \frac{d V}{d \phi} \,, \hspace{1cm}
\Gamma \equiv 
\frac{V (d^2V/d\phi^2)}{(dV/d\phi)^2} \, .
\end{eqnarray}
which contain all the relevant information about the potential.
Note that both $\lambda$ and $\Gamma$ are in general $\phi$ (and, thus, time)
dependent.

The exponential case, $V\sim e^{-\beta \kappa \phi}$,
is characterised by constant parameters ($\lambda = \beta $ and $\Gamma =1$);
the inverse power law potential, $V \sim \phi^{-\alpha}$, has constant 
$\Gamma =1 +1/\alpha$ and field--dependent $\lambda =-\alpha/\kappa\phi$.
More general potentials can give more complicated answers. 
As an example, the sugra-inspired class of potentials \cite{cnr} described by
$V \sim e^{\alpha (\kappa\phi)^{\beta}} (\kappa\phi)^{\mu}$ gives
\begin{eqnarray}
\label{lagama}
\lambda = - \frac{\alpha \beta (\kappa \phi)^{\beta} + \mu }{\kappa \phi} \, ,
\hspace{0.8cm}
\Gamma -1 = \frac{\alpha \beta (\beta-1) (\kappa \phi)^{\beta} -\mu }
                 {(\alpha \beta (\kappa \phi)^{\beta} + \mu)^2} \, .
\end{eqnarray}
\section{The attractor solutions}

As illustrated in Fig.~1 and Fig.~2, for a generic scalar potential one can 
identify up to five regions in phase space.
In these figures, region 1 represents a regime in which the
potential energy rapidly converts into kinetic energy; in region 2,
the kinetic energy is the dominant contribution to the total energy
density of the scalar field (``kination''); in region 3, the field
remains nearly constant until the attractor
solution is reached (``frozen field''); in region 4 the field evolves 
along the attractor
solution, where the ratio of the kinetic to potential energy is 
constant or slowly varying; and in region 5 the potential energy
becomes important, the scalar field dominates and drives the
dynamics of the Universe.
In the following we will present the analytic solutions for regions 4 and 5.

\begin{figure}[ht!]
\centerline{\vbox{
\psfig{figure=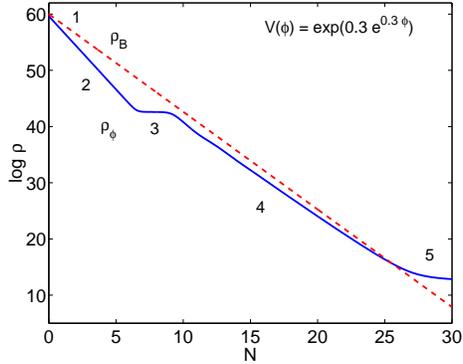,height=5cm}
}}
\caption[]{\label{fig1} Evolution of the scalar field energy density
$\rho_{\phi}$ in a radiation background fluid $\rho_B$ for a double exponential
potential. }
\end{figure}

\begin{figure}[ht!]
\centerline{\vbox{
\psfig{figure=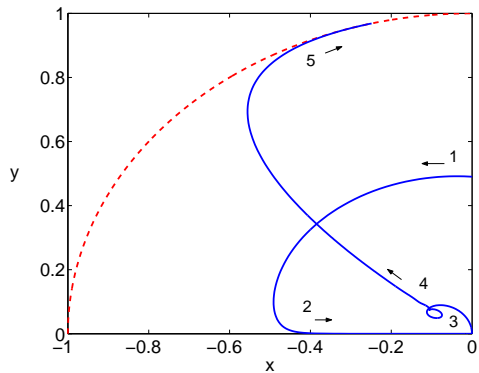,height=5cm}
}}
\caption[]{\label{fig2} The scalar field dynamics in the phase space
diagram for a double exponential potential. Regions 1 to 5 are the
same as in Fig. \ref{fig1}. The dashed line represents the unit
circle $x^2+y^2 = 1$.}
\end{figure}

%
%
{\bf The tracker solution.}
Consider the case in which the scalar field
evolution is well approximated by a linear relation between $x$ and $y$
(region 4 in the figures).
If we take the approximation of $\lambda$ large and $\Gamma$ 
nearly constant, then Eqs.~(\ref{dynamics}) give
the ``instantaneous critical points''
\begin{eqnarray}
\label{critical1}
x_c(\lambda) = \sqrt{\frac{3}{2}} \frac{\gamma_{\phi}}{\lambda} \,, \hspace{1cm}
y_c^2(\lambda) = \frac{3}{2} \frac{\gamma_{\phi}}{\lambda^2}(2-\gamma_{\phi}) \,, 
\end{eqnarray}
where the equation of state of the scalar field is
\begin{equation}
\label{eqstate}
\gamma_{\phi} = \frac{1}{2} [\gamma + (2\Gamma -1)\frac{\lambda^2}{3}] -
\frac{1}{2} \sqrt{ \left[ - \gamma + (2\Gamma-1)\frac{\lambda^2}{3} \right]^2 
+ 8\gamma(\Gamma-1)\frac{\lambda^2}{3} } \,.
\end{equation}
This solution exists for $\lambda^2 > 3\gamma_\phi$.

The scalar contribution to the total
energy density is $\Omega_{\phi} = 3 \gamma_{\phi}/\lambda^2$.
Note that when $\Gamma-1 \approx 0$ one has
\begin{equation}
\label{eqstate2}
\gamma_{\phi} = \gamma \left[1- \frac{2(\Gamma-1)}{1 
                                   -3\gamma/\lambda^2}   \right]   \,.
\end{equation}
In other words, for potentials with small curvature, the equation of
state of the scalar field is very close to the equation of state of
the background fluid, and it is said that the field ``tracks'' the
background.
This expression can account for values of $\Omega_{\phi} > 1/2$.

From Eq. (\ref{eqstate}), it can also be shown that in the limit of
$\lambda \rightarrow \infty$,
when the background fluid is completely dominating,
we recover the expression in \cite{stein},
\begin{equation}
\label{steinhardt}
\gamma_{\phi} = \frac{\gamma}{2\Gamma-1} .
\end{equation}

{\bf The scalar field dominated solution.}
Consider now the case in which the evolution of the scalar field 
approaches the unit circle $x^2+y^2 = 1$ (region 5 in the figures).                
One can note that at late times $y \sim 1$ and $x \rightarrow 0$. 
This means that the scalar potential is overtaking the energy density and 
the potential is very flat. 
In other words, $\lambda$ is getting closer to zero. Let us take then
$ \lambda \approx 0$ and $\lambda^2(\Gamma-1)$ nearly constant.
With these assumptions, we find critical points in,
\begin{eqnarray}
\label{critical2}
x_c(\lambda) = \frac{\lambda_{\phi}}{\sqrt{6}} \,, \hspace{1cm}
y_c^2(\lambda) = 1 - \frac{\lambda_{\phi}^2}{6}\,.
\end{eqnarray}
Hence, the scalar field is dominant, $\Omega_{\phi} = 1$ and 
$\gamma_{\phi} = \lambda_{\phi}^2/3$, where
we have defined
\begin{equation}
\label{lambdaphi}
\lambda_{\phi} = \frac{3}{2} \left[\frac{1 
                - \sqrt{1-4(\Gamma-1)\lambda^2/3}}
                                     {(\Gamma-1)\lambda}\right] , 
\end{equation}
for $\Gamma \neq 1$, and $\lambda_{\phi} = \lambda$ otherwise.
Expanding $\lambda_{\phi}$ we can approximate the solution by
\begin{equation}
\label{lfi}                                      
\lambda_{\phi} = \lambda \left[1+\frac{1}{3}(\Gamma-1)\lambda^2 \right] \,. 
\end{equation}
This solution exists for $\lambda_\phi^2 < 6$.

\vskip0.2cm

We would like to stress that all the results presented in this talk 
are completely general, since no assumption on the shape of the potential 
was made in the calculation.
A detailed discussion of the existence conditions and stability of the
two attractors can be found in \cite{nnr}, together with an illustration of
possible cosmological applications.

\acknowledgements{It is a pleasure to thank N.~J.~Nunes and S.~C.~C.~Ng 
with whom the work presented here was done.}

\vfill
\end{document}